\documentclass[%
aip,
jcp,
 amsmath,amssymb,
 reprint,%
]{revtex4-1}

\usepackage{graphicx}
\usepackage{dcolumn}
\usepackage{bm}
\usepackage{tabularx}
\usepackage{siunitx}
\usepackage{hyperref}
\hypersetup{
    colorlinks=true,
    linkcolor=blue,
    filecolor=magenta,
    citecolor=blue,
    urlcolor=blue,
    }
\usepackage{cleveref}
\DeclareSIUnit\angstrom{\text {Å}}

\usepackage[utf8]{inputenc}
\usepackage[T1]{fontenc}
\usepackage{mathptmx}
\usepackage{etoolbox}
\usepackage{bibentry}
\usepackage{hhline}

\makeatletter
\def\@email#1#2{%
 \endgroup
 \patchcmd{\titleblock@produce}
  {\frontmatter@RRAPformat}
  {\frontmatter@RRAPformat{\produce@RRAP{*#1\href{mailto:#2}{#2}}}\frontmatter@RRAPformat}
  {}{}
}%
\makeatother

\usepackage[dvipsnames]{xcolor}
\usepackage[normalem]{ulem}
\newcommand{\replace}[2]{#2}
\newcommand{\proof}[2]{#2}

\begin{document}

\preprint{AIP/123-QED}

\title[Variational Umbrella Seeding]{Variational Umbrella Seeding for Calculating Nucleation Barriers}
\author{Willem Gispen}
\affiliation{ 
Soft Condensed Matter \& Biophysics, Debye Institute for Nanomaterials Science, Utrecht University, Princetonplein 1, 3584 CC Utrecht, Netherlands 
}%
\author{Jorge R. Espinosa}
\author{Eduardo Sanz}
\author{Carlos Vega}%
\affiliation{ 
Departamento de Química Física, Facultad de Ciencias Químicas, Universidad Complutense de Madrid,
28040 Madrid, Spain}%
\author{Marjolein Dijkstra}%
 \email{m.dijkstra@uu.nl}
\affiliation{ 
Soft Condensed Matter \& Biophysics, Debye Institute for Nanomaterials Science, Utrecht University, Princetonplein 1, 3584 CC Utrecht, Netherlands 
}%


\date{\today}

\begin{abstract}

In this work, we  introduce Variational Umbrella Seeding, a novel technique for computing nucleation barriers. This new method, a refinement of the original seeding approach, is \replace{independent on}{far less sensitive to} the choice of order parameter for measuring the size of a nucleus. Consequently, it surpasses seeding in accuracy, and Umbrella Sampling in computational speed. We test the method extensively and demonstrate excellent accuracy for  crystal nucleation of nearly hard spheres and of two distinct models of water: mW and TIP4P/ICE. This method can easily be extended to calculate nucleation barriers for homogeneous melting, condensation, and cavitation.
\end{abstract}

\maketitle

A first-order phase transition from a metastable phase to a more stable phase often  occurs through the formation of a nucleus of the new phase. This nucleation process is apparent in phase transitions such as crystallization, melting, condensation, and cavitation. In crystallization, for example, nucleation determines the crystal polymorph and thus the material properties of the crystalline phase. More generally, the time required for the phase transition to start is determined by the nucleation rate. Simulations enable us to gain insight in the nucleation mechanisms and  the factors influencing the nucleation rate. This insight can then be used to understand, predict, or even control nucleation in natural and industrial processes.

However, nucleation is a rare event, meaning that nucleation takes a very long time  to occur spontaneously in simulations. 
Except for the simplest model systems, direct measurements of nucleation rates are thus not feasible in simulations. To overcome this rare-event challenge, several enhanced sampling techniques have been introduced, such as forward flux sampling,\cite{allen_sampling_2005} transition path sampling,\replace{}{\cite{bolhuis_transition_2002,lechner2011role,beckham2011optimizing}} metadynamics,\cite{laio_escaping_2002,trudu_freezing_2006} lattice mold,\cite{respinosa_lattice_2016} and Umbrella Sampling.\cite{auer_prediction_2001,torrie_nonphysical_1977} Although these techniques are much faster than brute-force simulations, they still require substantial computational resources. As a consequence, most nucleation studies focus on only  a few selected state points.

To gain  more insight into nucleation trends, it is useful to combine simulations with classical nucleation theory (CNT).\cite{volmer_keimbildung_1926,becker_kinetische_1935} The seeding technique\cite{espinosa_seeding_2016} leverages this approach to estimate  nucleation rates with much greater efficiency, and across wider ranges of conditions compared to traditional methods. This advancement has allowed the study of the thermodynamics of curved interfaces,\cite{montero_de_hijes_thermodynamics_2022} polymorph selection,\cite{gispen2023crystal} and nucleation phase diagrams.\cite{gispen2022kinetic,sadigh2021metastable} Despite its approximate nature,  the seeding technique  has \replace{}{been} demonstrated to \replace{give accurate results}{accurately capture the trends of nucleation rates} across a wide variety  of scenarios such as Lennard-Jones condensation, crystallization of hard spheres, water, and NaCl,\cite{espinosa_seeding_2016,fiorucci_effect_2020,dasgupta2020tuning,coli2021artificial} and melting of hard spheres.~\cite{gispen2024finding}


However, challenges arise in applying the seeding technique. At its core, this  technique relies on CNT approximations for the nucleation rate $J$ and the nucleation barrier $\Delta G^c$, given by
\begin{align}
    \label{eq:cnt-J}
    J &= J_0 \exp (- \Delta G^c_{\mathrm{CNT}} / k_B T), \\
        \label{eq:cnt-n}
    \Delta G^c_{\mathrm{CNT}} &= \frac{1}{2} n_c |\Delta \mu| .
\end{align}
Here $J_0$ denotes the kinetic prefactor, $k_B$ represents Boltzmann's constant, $T$  the temperature, $n_c$  the size of the critical nucleus, 
and $|\Delta \mu|$ is the supersaturation, i.e.\ the difference in chemical potential between the solid and liquid phases.
From the CNT approximation, \Cref{eq:cnt-n}, it is evident  that the nucleation barrier is directly determined  by the critical nucleus size $n_c$. To calculate this size, an order parameter (or criterion) is required to distinguish between the nucleus and  the surrounding parent phase. Different order parameter choices yield different nucleus sizes, and consequently,  different nucleation barriers. Thus the results of seeding simulations depend on the order parameter choice.  In contrast, \replace{}{the barrier heights obtained with} more rigorous rare-event methods such as Umbrella Sampling and metadynamics have been shown to \replace{depend only weakly}{depend less sensitively} on the chosen criterion for  measuring the nucleus size.\cite{filion_crystal_2010,prestipino2018barrier}


In this Article, we introduce a computational method that addresses the limitations  of the seeding technique. Like seeding, it relies on classical nucleation theory to  estimate  the nucleation barrier. However, it borrows ideas from Umbrella Sampling to improve this estimate. This combination results in Variational Umbrella Seeding, a method for estimating the barrier that is faster than Umbrella Sampling, while being more {accurate} than seeding.


\section{Seeding and Umbrella Sampling}

\begin{figure*}[ht]
    \centering
    \includegraphics[width=\linewidth]{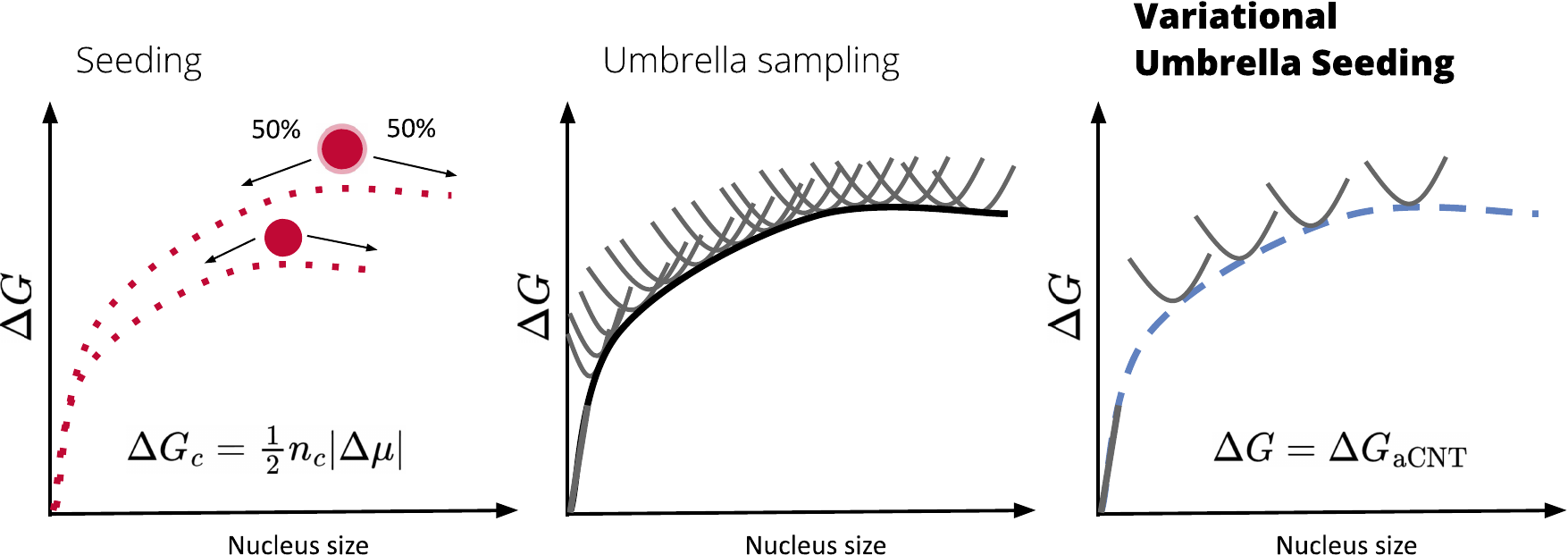}
    \caption{Schematic comparison of Seeding, Umbrella Sampling, and Variational Umbrella Seeding, illustrating how they estimate the nucleation barrier $\Delta G$ as a function of nucleus size. Seeding  employs the CNT approximation, $\Delta G^c = n_c|\Delta\mu|/2$, based on the number of particles $n_c$  in the critical nucleus for estimating the barrier height. This makes Seeding \replace{}{significantly} dependent on the choice of order parameter for computing  the nucleus size. In contrast, Umbrella Sampling and Variational Umbrella Seeding are \replace{independent of}{less sensitive to} the choice of order parameter. Umbrella Sampling does not rely on CNT and uses a large number of biased simulations (grey parabolas) to construct the nucleation barrier. Variational Umbrella Seeding relies on an adjusted version of CNT (aCNT) and uses a small number of biased simulations to construct the barrier.}
    \label{fig:method}
\end{figure*}

Before explaining our new method, we will first review two well-established popular methods for calculating nucleation barriers: Seeding\cite{espinosa_seeding_2016} and Umbrella Sampling.\cite{auer_prediction_2001} In \Cref{fig:method}, we show schematic representations of Seeding and Umbrella Sampling. 

In the seeding approach, the initial configuration consists of a nucleus - the `seed' - which is embedded within the metastable parent phase. Subsequently, simulations are performed to determine whether this seed grows or shrinks. By varying the temperature or the pressure, seeding identifies  the conditions under which the seed grows and shrinks with equal probability. This seed defines the critical nucleus under these conditions. Using the CNT approximation, \Cref{eq:cnt-n}, we can calculate the height of the nucleation barrier $\Delta G^c_{\mathrm{CNT}} = n_c |\Delta \mu|/2$.
This approximation has been \replace{shown to give accurate}{demonstrated to yield results for the nucleation barrier and nucleation rate that are consistent with more rigorous methods such as forward flux sampling and Umbrella Sampling, and even with completely unbiased `brute-force' simulations} for condensation, freezing, and melting of various systems such as Lennard-Jones, hard spheres, water, and NaCl.\cite{espinosa_seeding_2016,fiorucci_effect_2020,dasgupta2020tuning,coli2021artificial,gispen2024finding,montero_de_hijes_thermodynamics_2022,gispen2023crystal,gispen2022kinetic,sadigh2021metastable} \replace{}{With a reasonable choice of order parameter, the typical error of seeding is around 3-5 orders of magnitude in the nucleation rate. } Moreover, by relying on CNT, seeding requires only one piece of information from simulation: the critical nucleus size. This fact makes \replace{it very efficient and}{ Seeding very efficient since it can provide reasonably accurate estimates of the nucleation rate for wide ranges of conditions while being simple to implement and computationally inexpensive.}

\replace{}{Seeding efficiently identifies a nucleus that grows or shrinks with equal probability, unequivocally establishing its critical nature.}
However, the CNT approximation, \Cref{eq:cnt-n}, depends on the criterion \replace{chosen}{selected} to measure the \replace{nucleus size}{size of this critical nucleus}. 
In \Cref{fig:method}, the two red dots represent the \emph{same} critical nucleus measured with two different criteria. A looser criterion will classify more particles as part of the nucleus. This is visualized by the light red layer of particles on the surface of the nucleus. Consequently, employing a looser criterion, results in a larger critical nucleus size, and therefore a higher barrier height.
In summary, using two different criteria will yield estimates of  different critical nucleus sizes, and as a result different barrier heights. 

In contrast, Umbrella Sampling does not rely on CNT and depends \replace{only marginally}{far less sensitively} on the chosen criterion for measuring the  nucleus size.\cite{filion_crystal_2010,prestipino2018barrier} \replace{}{Note that Umbrella Sampling is not completely independent of this choice, a topic that we discuss further in \Cref{sec:comparison}.} Umbrella Sampling uses a large number of biased simulations to measure the free-energy profile. Typically, parabolic bias potentials, depicted as grey overlapping parabolas in \Cref{fig:method}, are used. Each parabola represents a separate biased simulation, where the bias potential serves to constrain the nucleus size. Essentially, \replace{by measuring the gradient along the entire free-energy profile, Umbrella Sampling effectively integrates the nucleation barrier in incremental steps.}{the biased simulations yield a large number of barrier segments, which are subsequently `stitched' together to construct the complete free-energy profile.} 
Although Umbrella Sampling does not rely on CNT, the requirement to use a  substantial number of biased simulations makes Umbrella Sampling considerably slower than seeding.

The idea of Variational Umbrella Seeding is to combine the strengths of Seeding \replace{as well as of}{and} Umbrella Sampling. By using an adjusted classical nucleation theory (aCNT), this approach {eliminates} the dependence of Seeding on the criterion chosen for measuring the nucleus size. At the same time, using aCNT also significantly reduces the number of biased simulations that are required to compute the barrier. In \Cref{fig:method}, we illustrate this by drawing only four parabolas instead of the many needed for Umbrella Sampling. 


\section{Adjusted Classical Nucleation Theory}
\label{sec:acnt}


Before introducing the adjusted classical nucleation theory (aCNT), we take a step back to the origin of the CNT expression, \Cref{eq:cnt-n}. 
For the purpose of generality, we consider a generic scenario where a nucleus of a `child' phase forms from a `parent' phase in the grand canonical ($\mu V T$)  ensemble.
According to CNT, the nucleation barrier is determined by two competing factors: the volume $v$ of the nucleus and the interface separating the nucleus from the parent phase. Without assuming a specific shape of the nucleus, we can write the area of this interface as $\replace{\alpha}{\alpha'} v^{2/3}$, where $\replace{\alpha}{\alpha'}$ is a dimensionless proportionality constant. Since $v^{1/3}$ is proportional to the linear dimension of the nucleus, it follows that the interfacial area is proportional to the square of the linear dimension.\cite{sharma2018nucleus} The grand potential of a nucleus with volume $v$ is then given by 
\begin{equation}
\label{eq:cnt-0}
    \Delta \Omega_{\mathrm{CNT}} (v) = \replace{\alpha}{\alpha'} v^{2/3} \gamma - v |\Delta P|,
\end{equation}
where $\gamma$ is the interfacial tension and $\Delta P$ is the Laplace pressure, i.e.\ the pressure difference between the child and parent phase at equal chemical potential. Maximizing $\Delta \Omega_{\mathrm{CNT}} (v)$ with respect to $v$ yields the critical nucleus size $v_c$, which  is related to the interfacial tension $\gamma$ as
\begin{equation}
    \label{eq:r*-0}
    2 \replace{\alpha}{\alpha'}\gamma =  {3 v_c^{1/3}} |\Delta P|.
\end{equation}
For a spherical nucleus of radius $r$, this becomes the familiar Laplace equation, $|\Delta P| = 2 \gamma / r$.
Substituting \Cref{eq:r*-0} into \Cref{eq:cnt-0} yields the  barrier height
\begin{equation}
    \label{eq:cnt-G-gamma-0}
    \Delta \Omega^c_{\mathrm{CNT}} = \frac{4 (\replace{\alpha}{\alpha'}\gamma)^3}{27 |\Delta P|^2  } = \frac{1}{2} v_c |\Delta P|.
\end{equation}

Now, the underlying approximations of \Cref{eq:cnt-n} become clearer. Essentially, the interfacial tension is determined by the critical nucleus size via \Cref{eq:r*-0}. This implies that the Laplace equation is implicitly contained in \Cref{eq:cnt-G-gamma-0}. Therefore, using \Cref{eq:cnt-G-gamma-0}, we assume that the criterion chosen for measuring the size of the nucleus corresponds to the interface of tension.\cite{montero_de_hijes_thermodynamics_2022} If this assumption does  not hold,  \Cref{eq:r*-0} will give an incorrect  value for the interfacial tension and therefore  an inaccurate value for the height of the barrier. \replace{}{From this discussion, it is also evident that the CNT approximation, \Cref{eq:cnt-n}, is not reliant  on any specific assumptions about the shape of the nucleus.\cite{sharma2018nucleus}}

In simulations of nucleation, it is more common to measure the Gibbs free-energy barrier $\Delta G$ in the isobaric-isothermal ($NPT$) ensemble. As Ref.\ \citenum{oxtoby1988nonclassical} discusses, the free-energy difference with respect to the homogeneous parent phase remains the same, i.e.\ $\Delta\Omega = \Delta G$. Therefore, the Gibbs free-energy barrier is given by 
\begin{subequations}
    \begin{equation} \label{eq:cnt}
        \Delta G_{\mathrm{CNT}}(v) = \replace{\alpha}{\alpha'} v^{2/3} \gamma - v |\Delta P|,
    \end{equation}
    \begin{equation} \label{eq:r*}
    2 \replace{\alpha}{\alpha'}\gamma =  {3 v_c^{1/3}} |\Delta P|,
    \end{equation}
    \begin{equation} \label{eq:cnt-G-gamma}
        \Delta G^c_{\mathrm{CNT}} = \frac{4 (\replace{\alpha}{\alpha'}\gamma)^3}{27 |\Delta P|^2  } = \frac{1}{2} v_c |\Delta P|.
    \end{equation}
\end{subequations}
From now on, we always assume an isobaric-isothermal ($NPT$) ensemble, so we focus on the Gibbs free-energy barrier $\Delta G$ for nucleation.

In the context of crystal nucleation, 
the Gibbs free-energy barrier $\Delta G (n)$ is usually measured
 as a function of the number of particles $n$ in the nucleus at a fixed pressure. The CNT equations corresponding to \Cref{eq:cnt-G-gamma,eq:cnt,eq:r*} become as follows \cite{sharma2018nucleus}
\begin{subequations}
    \begin{equation} \label{eq:cnt-G-n}
        \Delta G_{\mathrm{CNT}} (n) = \alpha n^{2/3} \gamma - n |\Delta \mu|,
    \end{equation}
    \begin{equation} \label{eq:cnt-gamma-n}
        2 \alpha\gamma =  {3 n_c^{1/3}} |\Delta \mu|,
    \end{equation}
    \begin{equation} \label{eq:cnt-gamma-G}
        \Delta G^c_{\mathrm{CNT}} = \frac{4 (\alpha\gamma)^3}{27 |\Delta\mu|^2  } = \frac{1}{2} n_c |\Delta \mu|.
    \end{equation}
\end{subequations}
In these equations, $|\Delta\mu|$ is the difference in chemical potential between the solid and liquid phase at equal pressure and temperature, and $n_c$ denotes the number of particles in the critical nucleus. The proportionality constant $\alpha$ is now expressed in units of area. Again, it is important to note that the criterion used  to measure  the nucleus size $n$  will influence the value of the interfacial tension $\gamma$ and therefore the height of the barrier.

\begin{figure*} 
    \centering
    \includegraphics[width=\linewidth]{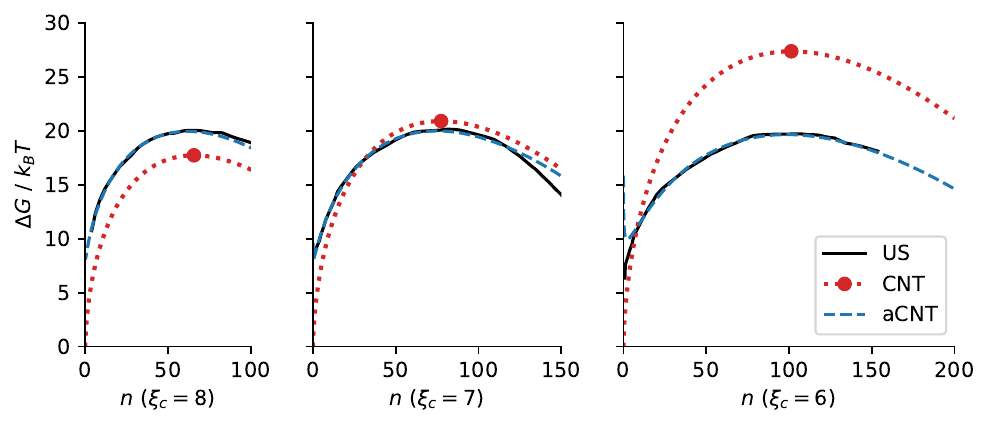}
    \caption{The Gibbs free energy  $\Delta G(n) /k_B T$ for the formation of a crystal nucleus of size $n$ from a fluid of hard spheres at a pressure of $P \sigma^3/k_B T = 17.0$. The black lines represent Umbrella Sampling results from Ref.  \citenum{filion_crystal_2010}. The blue lines denote  aCNT fits of the nucleation barriers, while the red dots and red dotted lines are CNT approximations of the barrier.}
    \label{fig:hs}
\end{figure*}

To illustrate this issue, we show in \Cref{fig:hs} Umbrella Sampling simulations  of hard-sphere nucleation as described in Ref. \citenum{filion_crystal_2010}. These crystal nucleation barriers were measured at a pressure $P \sigma^3/k_B T = 17$, corresponding to a metastable fluid packing fraction of $\eta = 0.5352$, a supersaturation of $|\Delta\mu| =0.54 ~k_B T$,  and a barrier height  $\Delta G^c = 19.7 \pm0.3 ~k_B T$. We show three different barriers, each measured with a different criterion to distinguish the crystal nucleus from the surrounding fluid. \replace{To be precise, the number of solid-like bonds $\xi_c$ used to classify a  particle as solid-like was varied.}{
To be precise, the  criteria used in Ref.\ \citenum{filion_crystal_2010} were as follows. First, a spherical harmonics expansion, $q_{lm}$, of the nearest neighbor density was computed, employing  a cutoff distance of $1.3\sigma$ (for $\xi_c=6, 8$) or $1.4\sigma$ (for $\xi_c=7$) to identify nearest neighbors. Subsequently, solid-like bonds were identified by assessing the  inner products, $d_6(i,j)$, between the $q_{6m}$ vectors of neighboring particles.  A pair of particles $(i,j)$ were considered to have a solid-like bond if  $d_6(i,j)>0.7$. A particle was classified as solid-like if it possessed $\xi_c$ or more solid-like bonds. Six different thresholds $\xi_c$ were investigated, of which we show $\xi_c=6, 7, 8$ for illustration. Two solid-like particles belong to the same cluster if their distance is less than $1.3\sigma$ (for $\xi_c=6, 8$) or $1.4\sigma$ (for $\xi_c=7$). Finally, the largest cluster of solid-like particles was identified as the crystal nucleus.}

Remarkably, the barrier height $\Delta G^c$  measured via Umbrella Sampling is almost independent of the criterion. In contrast, the critical nucleus size $n_c$ varies significantly. For $\xi_c=8$, the critical nucleus size is $n_c = 72$, while for $\xi_c=6$, it is $n_c=102$. Substituting these values into \Cref{eq:cnt-gamma-G}, we obtain nucleation barriers varying from $19 ~k_B T$ to $28 ~k_B T$.  We have plotted these approximations as red dots in \Cref{fig:hs}. In this case $\xi_c=7$ shows the best agreement with Umbrella Sampling. For $\xi_c=8$, the barrier is underestimated due to a too stringent  criterion, whereas for $\xi_c=6$, the barrier is overestimated because the criterion is too loose.

Apart from examining the height of the barrier, we can also investigate the shape of the barrier. By leveraging the critical nucleus size $n_c$, we can estimate the interfacial tension $\gamma$ using \Cref{eq:cnt-gamma-n}. In \Cref{fig:hs}, we  plot the CNT nucleation barrier,  \Cref{eq:cnt-G-n}, using this approximation for the interfacial tension as a red dotted line. We see that these representations deviate significantly from the Umbrella Sampling results denoted in black. Even in the best case of $\xi_c=7$, we see that the CNT expression for the barrier fails to accurately capture the shape of the barrier.
In short, this example of hard-sphere nucleation demonstrates that with an ill-defined  nucleus size, the CNT equations do not capture the height and shape of the nucleation barrier.

In the literature, numerous efforts have been made to refine CNT to offer a more precise description of the shape of the nucleation barrier. For instance, \citet{merikanto2007origin} observed deviations in the initial part of the barrier for  condensation. They proposed that this effect could be accounted for by a vertical shift of the free-energy barrier, essentially an additive term in \Cref{eq:cnt-G-n}. Additionally, \citet{prestipino_systematic_2012} introduced corrections to accommodate  a non-sharp and thermally fluctuating interface. Firstly, they added  a logarithmic term proportional to $\log (r)$, where $r$ denotes the radius of the nucleus. Secondly, they replaced the constant interfacial tension $\gamma$ with a function $\gamma(r) = \gamma_0 ( 1- 2\delta/r + \epsilon/r^2)$, in essence representing  curvature corrections to the interfacial tension. They noted that the logarithmic correction has a considerably  smaller effect on the quality of the fit than the curvature corrections to the interfacial tension.
Finally, \citet{filion_crystal_2010} suggested that the effect of different order parameters can be captured by a shift in the radius. They proposed to replace the measured radius $r$  by $r-\delta$ in the CNT equations.

Rather than selecting one of these modifications, we propose the following expressions for the shape of the nucleation barrier
\begin{subequations}
\label{eq:acnt}
\begin{alignat}{2}
\label{eq:acnt1}
    \Delta G_{\mathrm{aCNT}} (v) &= - v |\Delta P| + g'_2 v^{2/3} + g'_1 v^{1/3} + g'_0,\\
\label{eq:acnt2}
    \Delta G_{\mathrm{aCNT}} (n) &= - n |\Delta \mu| + g_2 n^{2/3} + g_1 n^{1/3} + g_0.
\end{alignat}
\end{subequations}
In these expressions, $g'_i$ and $g_i$ are fitting parameters for the nucleation barrier. 
Following Ref.~\citenum{filion_crystal_2010}, we refer to \Cref{eq:acnt1,eq:acnt2} as the adjusted classical nucleation theory (aCNT) expressions for the nucleation barrier. \replace{}{Given that  there are three free parameters, these equations could alternatively  be referred to as ``three parameter CNT'' expressions.} In contrast to Refs. \citenum{merikanto2007origin,prestipino_systematic_2012,filion_crystal_2010}, we do not assign a specific physical interpretation to these parameters. However, we note that all their corrections, except for the logarithmic term, are captured by these aCNT expressions.


To validate the aCNT expressions, we use \Cref{eq:acnt2} to fit the Umbrella Sampling barriers of hard-sphere crystal nucleation. We show these fits as dashed blue lines in \Cref{fig:hs}. During the fitting procedure, we kept the supersaturation $|\Delta\mu|=0.54~k_B T$ fixed, as determined by thermodynamic integration of the equation of state. The remaining three parameters $g_0, g_1, g_2$ are fitted to the Umbrella Sampling data. More specifically, they were fitted to 
the part of the nucleation barrier where the nucleus size $n$ obeys $5 < n < n_c$. In this regime, the barrier as measured by Umbrella Sampling deviates less than $0.4 ~k_B T$ from the aCNT fit. From this comparison, it is evident  that the aCNT expression effectively captures the shape of the nucleation barrier of hard spheres, as measured with different order parameters.

\begin{figure*}
    \centering
    \includegraphics[width=\linewidth]{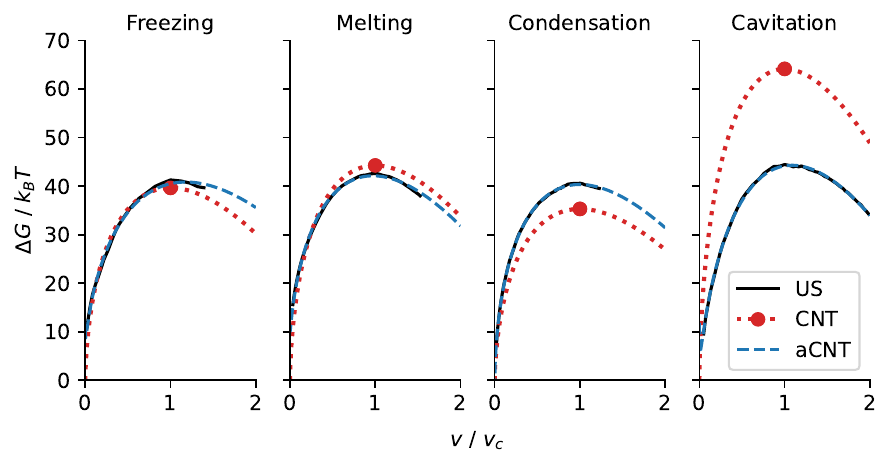}
    \caption{The Gibbs free energy  $\Delta G(v)/k_B T$ for nucleation during freezing~\cite{filion_crystal_2010} and  melting~\cite{gispen2024finding} of hard spheres, condensation of Lennard-Jones,\cite{sanchez-burgos_equivalence_2020} and cavitation of water.\cite{menzl_molecular_2016} The size of the nucleus $v$ is normalized by the critical nucleus size $v_c$. The solid black lines are the Umbrella Sampling measurements of Refs. \citenum{filion_crystal_2010,gispen2024finding,sanchez-burgos_equivalence_2020,menzl_molecular_2016}. The dashed blue lines denote aCNT fits of the nucleation barriers, while the red dots and red dotted lines are CNT approximations of the barrier.}
    \label{fig:four-phase-transitions}
\end{figure*}

As a further illustration, we investigate four different nucleation barriers previously measured using Umbrella Sampling for various phase transitions: crystal nucleation,\cite{filion_crystal_2010} crystal melting,\cite{gispen2024finding} cavitation,\cite{menzl_molecular_2016} and condensation.\cite{sanchez-burgos_equivalence_2020} In \Cref{fig:four-phase-transitions}, the  measured barriers are denoted as solid black lines. The  CNT approximations of these barriers,  \Cref{eq:cnt,eq:r*,eq:cnt-G-gamma}, based on the volume $v_c$ of the critical nucleus are represented as red dots and red dotted lines. We observe  that the freezing and melting barriers are reasonably well described  by the CNT approximation. However, the  condensation barrier is underestimated by CNT, while  the cavitation barrier is severely overestimated. It is important to stress again that the accuracy of the CNT approximation depends critically on the criterion used to identify the nucleus. For instance, for condensation and cavitation, the `equidensity' criterion has been shown to be a reasonable choice.\cite{sanchez-burgos_equivalence_2020} Next, we employ the aCNT expression, \Cref{eq:acnt1}, to fit the barriers. In this case, the Laplace pressure $|\Delta P|$ is determined by thermodynamic integration of the equation of state, while the remaining three parameters $g'_0, g'_1, g'_2$ are fitted to the Umbrella Sampling data. The resulting fits are shown with dashed blue lines. \replace{It is evident that the shape of all these nucleation barriers can be well described  with aCNT.}{
We note that not any functional form with three free parameters will give such good fits as the aCNT expression. For example, if we use a quadratic fitting function in the nucleus volume $v$ or a quadratic function in $v^{1/3}$, we get worse fits of the barriers in Figure 3. Overall, it is evident that the shape of all these nucleation barriers can be well described with aCNT.}

\section{Method}


The effectiveness of aCNT in describing the nucleation barrier of crystallization, cavitation, condensation, and melting, suggests a strategy for approximating the nucleation barrier height: By fitting the parameters $g_0,g_1,g_2$ in \Cref{eq:acnt}, the height of the barrier can be determined  without the need of performing full Umbrella Sampling calculations. In this Section, we will introduce Variational Umbrella Seeding as an efficient implementation of this idea.

\begin{figure*}[!htbp]
    \centering
    \includegraphics[width=\linewidth]{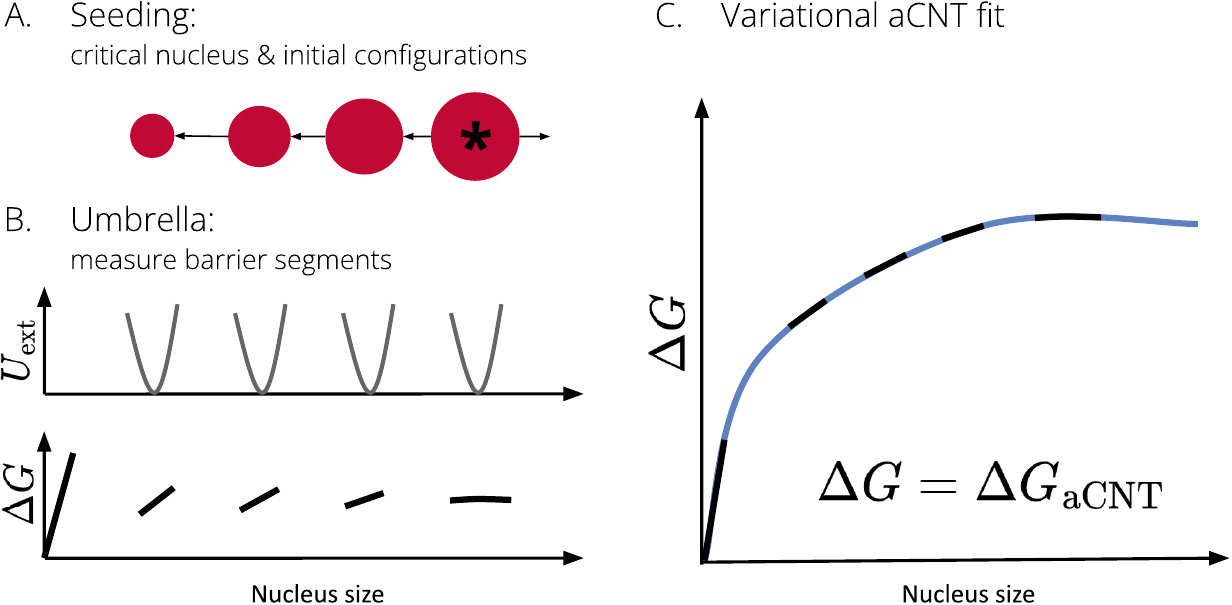}
    \caption{Schematic overview of Variational Umbrella Seeding. A. Initially,  Seeding is employed to estimate  the critical nucleus size and to provide initial configurations for the umbrella sampling runs. B. Subsequently, Umbrella Sampling is used to measure segments of the nucleation barrier, along with unbiased simulations to measure the initial part of the barrier. C. Finally,  an adjusted version of CNT (aCNT) is employed to construct the entire nucleation barrier from these barrier fragments.}
    \label{fig:vus-method}
\end{figure*}


Variational Umbrella Seeding combines ideas from both Seeding and Umbrella Sampling. In \Cref{fig:vus-method}, we provide a schematic overview of the method. It employs Seeding to offer a first estimate of the critical nucleus size and to provide initial configurations for the umbrella sampling runs. Starting from these initial configurations, we use Umbrella Sampling to measure segments of the nucleation barrier. Subsequently, these barrier segments are fitted to an aCNT expression to approximate  the full nucleation barrier. 


\subsection{{Seeding: critical nucleus and initial configurations}}
The first step in Variational Umbrella Seeding is to determine the critical nucleus size and to generate some initial configurations. The Seeding technique is very efficient for this purpose. The result of the seeding technique is a configuration of a nucleus under the conditions for which this nucleus is critical. Hence, we already possess a good estimate of the critical nucleus size $n_c$. By starting simulations from the critical nucleus, we can select those simulations for which the nucleus shrinks. From these trajectories, we can then readily select initial configurations of nucleus sizes ranging from $0$ to $n_c$. \replace{}{We note that it is not strictly necessary to use Seeding for this first step if the critical nucleus size and initial configurations can be obtained by other means. However, Seeding is a very convenient and efficient method for this purpose.}

\subsection{Measure barrier segments with Umbrella Sampling}
\label{sec:umbrella}

The second step involves  measuring segments of the nucleation  barrier using Umbrella Sampling. As illustrated in \Cref{fig:vus-method}, only  four of these barrier fragments are required.

To perform these simulations, we employ the hybrid Monte Carlo method as a  popular way for performing Umbrella Sampling simulations of nucleation barriers.\cite{gonzalez2014nucleation,guo2018hybrid,guidarelli2023neural} The main advantage of hybrid Monte Carlo over traditional Monte Carlo is that it allows the use of highly optimized and parallelized molecular dynamics codes such as LAMMPS\cite{plimpton_fast_1995} and GROMACS.\cite{bekker1993gromacs} 

In essence, hybrid Monte Carlo combines  Monte Carlo simulation with molecular dynamics (MD) simulations to propose trial moves for Monte Carlo (MC). Each MC cycle consists of the following steps. First, initial velocities are generated for each particle according to a Maxwell-Boltzmann distribution. Subsequently, a short trajectory is simulated using MD. The trajectory is accepted or rejected according to a Metropolis criterion based on the total energy of the system. 

We constrain the nucleus size by adding an external bias potential to the total energy in the Metropolis acceptance step. To elaborate, after each MD trajectory, we accept the trajectory with a probability $p_{\mathrm{acc}}$ given by
\begin{align} 
    p_{\mathrm{acc}} &= \min \left(1, \exp(-\Delta H/k_B T) \right ), \label{eq:nve-accept} \\
    H &= K + U_{\mathrm{pot}} + U_{\mathrm{ext}} (n), \\
    U_{\mathrm{ext}} (n) &= \frac{1}{2} k (n - \hat{n})^2.
\end{align}
In these equations, $\Delta H$ represents the change in the total energy $H$ of the system, which consists of the kinetic energy $K$, potential energy $U_{\mathrm{pot}}=\sum_{i<j}\phi(r_{ij})$, the external bias potential $U_{\mathrm{ext}}$ and $\phi(r_{ij})$ the pair interaction between particle $i$ and $j$.
The external bias potential $U_{\mathrm{ext}}$ is employed  to keep the nucleus size $n$ close to the target size $\hat{n}$. The spring constant $k$ determines the stiffness of the bias potential. The result of such a simulation is a biased time series of nucleus sizes $n(t)$. This time series is subsequently converted into a biased nucleus size distribution $p(n)$. From this biased nucleus size distribution, we can estimate the unbiased  free-energy profile
\begin{equation} \label{eq:G-from-p}
    G(n)/k_B T = -\ln p(n) - U_{\mathrm{ext}}(n)/k_B T + \textrm{constant}.
\end{equation}

To sample the isobaric-isothermal ($NPT$) ensemble, we consider two approaches. The first approach is to use an $NVE$ integrator for the MD trajectories and perform volume moves after each trajectory. Trajectories are accepted according to \Cref{eq:nve-accept} and the volume moves according to standard volume move acceptance criteria, taking the bias potential into account.
By using a symplectic and time-reversible $NVE$ integrator, this scheme guarantees that detailed balance is obeyed.\cite{mehlig1992hybrid}

Alternatively, \citet{gonzalez2014nucleation} showed that an $NPT$ integrator can also be employed for the MD trajectories. In this case, no separate volume moves are performed. Trajectories are accepted with a probability
\begin{equation} \label{eq:npt-accept}
    p_{\mathrm{acc}} = \min \left(1, \exp(- \Delta U_{\mathrm{ext}}) /k_B T \right ).
\end{equation}
Although this scheme does not guarantee detailed balance, \citet{gonzalez2014nucleation} showed that this approach provides  an excellent approximation provided that the MD time step is small enough and the MD trajectories are long enough. 

To evaluate our hybrid Monte Carlo methods, we measured the density and potential energy of the homogeneous liquid phase without an external bias, i.e.\ $U_{\mathrm{ext}}=0$ in both \Cref{eq:nve-accept} and \Cref{eq:npt-accept}. In this way, we can compare the density and potential energy directly to a standard unbiased MD simulation, where  velocities are not resampled. In all cases, we made sure that the relative error in both the density and potential energy was less than $0.05 \%$.

Variational Umbrella Seeding relies on a small number of biased simulations. Similarly to Umbrella Sampling, these simulations should be evenly distributed across the nucleation barrier. We found that employing four different simulations, with target sizes $\hat{n} = \hat{n}_c/4$, $\hat{n}_c/2$, $3 \hat{n}_c/4$, and $\hat{n}_c$, yields reliable results. This means that  having a rough estimate of the critical nucleus size $\hat{n}_c$ before starting the simulations, is useful but this estimate does not need to be accurate. This estimate of the critical nucleus size  aids in selecting the spring constant $k$ of the bias potential. In \Cref{sec:idealized}, we describe our approach for  optimizing these parameters of Variational Umbrella Seeding using tests with an idealized nucleation barrier.

\subsection{Variational aCNT fit of the nucleation barrier}
A nucleation barrier is usually constructed from a sequence  of biased simulations by `stitching' together many overlapping free-energy \replace{profiles}{segments}. Common approaches for this task include the Multistate Bennett Acceptance Ratio (MBAR) method or the Weighted Histogram Analysis Method (WHAM).
In order to take advantage of the aCNT, \replace{\Cref{eq:acnt},}{} 
we adopt a different approach to process the simulation data. \replace{}{The key idea is to fit the parameters $g_0,g_1,g_2$ in \Cref{eq:acnt} directly from  simulation data. The height of the barrier can then be determined by extracting the maximum from the resulting aCNT fit. To perform this fit, we require a scheme that does not rely on overlapping free-energy segments and can estimate parameters based on an assumption for the functional form of the free-energy profile.} \replace{}{In this regard, we drew  inspiration from the Variational Free-Energy Profile (VFEP) \cite{lee2013new} method proposed within the context of biomolecular reactions. However, rather than employing a generic spline interpolation function, we take advantage of the aCNT Equation (8) to improve the efficiency of the method.} \replace{W}{In essence, w}e employ a maximum-likelihood approach to fit the parameters $g_1$ and $g_2$, while employing a separate unbiased simulation to fit $g_0$.

To illustrate the maximum-likelihood approach, we first compute the likelihood of a series of measurements, under the assumption that the aCNT \Cref{eq:acnt} holds for certain values of the parameters $g_i$. In this case, the nucleation barrier and the external bias potential $U_{\mathrm{ext}}$ generate a probability distribution $p(n)$ for the nucleus size, given by 
\begin{align}
    p(n) &= \frac{1}{Z}\exp[- (\Delta G_{\mathrm{aCNT}} (n) +  U_{\mathrm{ext}} (n)) /k_B T], \\
    Z    &= \sum_n \exp[- (\Delta G_{\mathrm{aCNT}} (n) +  U_{\mathrm{ext}} (n)) /k_B T].
\end{align}
We note that the normalization factor $Z$ depends on both the aCNT parameters $g_i$ and the bias potential $U_{\mathrm{ext}}$, but $Z$ can easily be computed. The likelihood $L$ of a series of independent measurements $n(t)$ is given by 
\begin{equation} \label{eq:L}
    L(n(t)) =  \prod_t \frac{1}{Z}\exp[- (\Delta G_{\mathrm{aCNT}} (n(t)) +  U_{\mathrm{ext}} (n(t))) /k_B T]
\end{equation}
where $t$ is the simulation time. With a logarithmic transformation, this becomes the log-likelihood $\ell$ given by
\begin{equation} \label{eq:logL}
    \ell(n(t)) = -\sum_t \left [\ln Z + (\Delta G_{\mathrm{aCNT}} (n(t)) +  U_{\mathrm{ext}} (n(t)))/k_B T \right ].
\end{equation}
Finally, the total log-likelihood of multiple simulations is
the sum of the individual log-likelihood functions
\begin{equation} \label{eq:logL-total}
    \ell (n_1(t), \dots) = \sum_i \ell_i (n_i(t)).
\end{equation}
Here $\ell_i$ represents the likelihood of the series of measurements $n_i(t)$ in simulation $i$. 
One subtlety in this derivation is that we assumed that the measurements $n(t)$ are independent of each other. However, the measurements $n(t)$ originate from a correlated time series, thus  this assumption does not automatically hold. To solve this issue, we estimate the autocorrelation time $\tau$ of the time series $n(t)$ as the time at which its normalized autocorrelation function has decayed to $1/e$. We employ  subsampling to obtain independent samples: rather than using the entire series $n(t)$ to compute the log-likelihood \Cref{eq:logL}, we only use a subset $t_1, t_2, \dots$ with a separation equal to the autocorrelation time i.e.\ $t_{i+1} = t_i + \tau$.

In short, for a given set of aCNT parameters $g_i$, we use \Cref{eq:logL-total} to compute the log-likelihood of our observations $n_i(t)$.  The maximum likelihood estimator for $g_1, g_2$ is consequently the set of parameters $\hat{g}_1,\hat{g}_2$ that maximizes the log-likelihood.
\replace{We note that this maximum-likelihood approach is very similar to the Variational Free-Energy Profile (VFEP) \cite{lee2013new} method, which  was proposed in the context of biomolecular reactions. However, instead of using a generic spline interpolation function, we take advantage of the aCNT \Cref{eq:acnt} to improve the efficiency of the method.}{}

Because $g_0$ is an additive term in the nucleation barrier, the log-likelihood \Cref{eq:logL-total}  is independent of $g_0$.
To fit $g_0$, 
we perform an independent and unbiased simulation of the metastable liquid. \replace{In this way, we also measure the full nucleus size distribution $p (n)$ of the metastable liquid, and we do not rely on aCNT to determine the shape of this initial part of the nucleation barrier.}{} The initial part of the nucleation barrier is then given by
\begin{equation} \label{eq:G-convention}
    \Delta G (n) / k_B T = - \ln \left(\frac{\replace{p(n)}{N_n}}{N} \right ).
\end{equation}
To be precise, when we refer to  $\replace{p(n)}{N_n}$, we denote the average number of nuclei of size $n$ in a system of $N$ particles. \replace{In this way,}{} \Cref{eq:G-convention} quantifies the free energy with the homogeneous liquid serving as the reference state.\cite{auer_numerical_2004}
Subsequently, the unbiased initial part of the barrier $\Delta G$ is `glued' to the aCNT barrier, and this `glueing'  process  determines the value of $g_0$. To establish this, we determine the nucleus size $n_0$ for which $\Delta G(n_0) \approx 10 ~k_B T$. 
Subsequently, we determine $g_0$ such that the equality
\begin{equation}
   \label{eq:glue-g0}
    \Delta G_{\mathrm{aCNT}} (n_0 | g_i) = \Delta G (n_0)
\end{equation}
is satisfied.
\replace{}{In this way, we do not rely on aCNT 
to provide an accurate free energy description of $\Delta G(n)$  for values of $n$ smaller than $n_0$ but rather for larger values which are the relevant ones for nucleation. Intuitively, $g_1$ and $g_2$ contain information about the gradient of the free energy profile (i.e.\ $d\Delta G(n)/dn$), whereas $g_0$ contains information about the absolute value of the free energy with respect to the homogeneous liquid.}

Finally, when we have fitted the aCNT parameters $g_i$, we obtain the Variational Umbrella Seeding estimate for the nucleation barrier height simply as the maximum of the aCNT \Cref{eq:acnt}. \replace{}{From the location of this maximum, we also estimate the critical nucleus size $n_c$.}


\subsection{Pressure/temperature dependence of the nucleation barrier}
\label{sec:temp-dG}
Up to this point, we have described the procedure for employing  Variational Umbrella Seeding to obtain an estimate of the nucleation barrier $\Delta G^c$ for a single state point. 
Once the nucleation barrier is estimated for several state points, we can fit them with a CNT expression to obtain the pressure or temperature dependence of the nucleation barrier. This process closely resembles that of the Seeding approach, with one notable exception: the approximation of the interfacial tension. From a  Variational Umbrella Seeding estimate $\Delta G^c_{\mathrm{aCNT}}$, we compute the interfacial tension using \Cref{eq:cnt-gamma-G} as follows
\begin{equation} \label{eq:acnt-gamma}
    \alpha\gamma = \left(\frac{27 \Delta G^c_{\mathrm{aCNT}} |\Delta\mu|^2}{4}\right)^{1/3}.
\end{equation}
\replace{}{Note that the proportionality constant $\alpha$ has units of area, so both sides of this equation have units of energy.}
In the Seeding approach, the interfacial tension is directly determined  by the critical nucleus size $n_c$. In contrast, here we determine the interfacial tension from the barrier height $\Delta G^c_{\mathrm{aCNT}}$. For a spherical nucleus, \replace{}{where $\alpha = \left ( 36 \pi / \rho_s^2 \right )^{1/3}$, }this can be expressed as
\begin{equation} \label{eq:acnt-gamma-spherical}
    \gamma = \left(\frac{3\Delta G^c_{\mathrm{aCNT}} |\Delta\mu|^2 \rho_s^2}{16 \pi}\right)^{1/3},
\end{equation}
where $\rho_s$ is the \replace{}{number} density of the solid phase.
Subsequently, similar to the Seeding approach, the interfacial tension $\gamma (P, T)$ is fitted  as a function of pressure or temperature. The fit is usually linear\cite{espinosa_seeding_2016}, but if the data suggests otherwise, one can fit to another functionality. The pressure and temperature dependence of the nucleation barrier is then given by
\begin{subequations} \label{eq:temp-dG}
    \begin{alignat}{2}
        \Delta G^c_{\mathrm{aCNT}} (P,T) &= \frac{4 (\alpha\gamma(P,T))^3}{27  |\Delta\mu(P,T)|^2},  \label{eq:temp-dG-a} \\
        \Delta G^c_{\mathrm{aCNT}} (P,T) &= \frac{16 \pi (\gamma(P,T))^3}{3 |\Delta\mu(P,T)|^2 \rho_s(P,T)^2}.\label{eq:temp-dG-b}
    \end{alignat}
\end{subequations}
Again, the second equation refers to the case of a spherical nucleus. We emphasize  that both \Cref{eq:temp-dG-a,eq:temp-dG-b} yield identical  results for $ \Delta G^c_{\mathrm{aCNT}} (P,T)$, irrespective of the actual shape of the nucleus. Imposing a specific nucleus shape  is only necessary for the interpretation of the interfacial tension.
To calculate the nucleation rate, we also require the kinetic prefactor. This is exactly the same as in Seeding\cite{espinosa_seeding_2016} and in Umbrella Sampling.\cite{auer_prediction_2001,auer_numerical_2004}







\section{Results and Discussion}

We assess the performance of  Variational Umbrella Seeding on crystal nucleation in three distinct systems: nearly hard spheres (WCA), monatomic water (mW), and the TIP4P/ICE model of water. 

\subsection{Nearly hard spheres (WCA)}
We simulate  hard spheres using  a Weeks-Chandler-Andersen (WCA) potential. This potential is the repulsive part of the Lennard-Jones pair potential and is widely used in molecular dynamics simulations to model hard spheres. In Ref.\ \citenum{filion_simulation_2011}, it was demonstrated that an effective hard-sphere diameter $\sigma_{\mathrm{eff}}=1.097\sigma$ effectively captures the free energy and nucleation rates of hard spheres. This effective hard-sphere diameter is defined such that the freezing density of the WCA system maps onto the freezing density of the hard-sphere system. Following Ref.\ \citenum{filion_simulation_2011}, we measure the nucleation barrier at a temperature of $k_B T / \epsilon = 0.025$ and a pressure of $P \sigma^3 / k_B T= 12.0$. We simulate \proof{$N=2\times 10^4$}{$N=8\times 10^3$} particles, employing a time step of $0.001 \tau_{\mathrm{MD}}$, where $\tau_{\mathrm{MD}}=\sqrt{m \sigma^2 / k_B T}$ represents the molecular dynamics unit of time. The temperature and pressure are maintained fixed with a Nos\'e-Hoover thermostat and barostat with relaxation times of $0.1 \tau_{\mathrm{MD}}$ and $0.5 \tau_{\mathrm{MD}}$, respectively. To measure the nucleus size, we  use the same order parameter as in Ref.\ \citenum{filion_simulation_2011}. To be precise, we calculate a spherical harmonics expansion $q_{lm}$ of the nearest neighbor density, where we use a cutoff of $1.5\sigma$ to identify nearest neighbors. Next, we compute inner products \replace{$q_l(i,j)$}{$d_6(i,j)$} between the \replace{$q_{lm}$}{$q_{6m}$} vectors of neighboring particles, and identify a solid-like bond as a pair of particles $(i,j)$ for which \replace{$q_l(i,j)>0.7$}{$d_6(i,j)>0.7$}. A particle is solid-like if it has $\xi_c$ or more solid-like bonds. We explore different values of $\xi_c = 6,7,8$ to investigate the effect  of different order parameters. \replace{}{Two solid-like particles belong to the same cluster if their distance is less than $1.5\sigma$.} Finally, the largest cluster of solid-like particles is identified as the crystal nucleus. Based on Ref.\ \citenum{filion_simulation_2011}, we use the following initial estimates for the critical nucleus sizes: $\hat{n}_c=185$, $155$, and $130$ for $\xi_c = 6,7,8$, respectively.
As described in \Cref{sec:umbrella}, these estimates also determine the target sizes and spring constants in the hybrid Monte Carlo simulations. In these simulations we use short molecular dynamics trajectories with a length of $0.5 \tau_{\mathrm{MD}}$ amounting to a total simulation time of $100,000 \tau_{\mathrm{MD}}$ for each target size.  We used the first $10,000 \tau_{\mathrm{MD}}$ for equilibration and the following $90,000 \tau_{\mathrm{MD}}$ for production. We found the autocorrelation time to be approximately $100 \tau_{\mathrm{MD}}$.

\begin{figure}[!htbp]
    \centering
    \includegraphics[width=\linewidth]{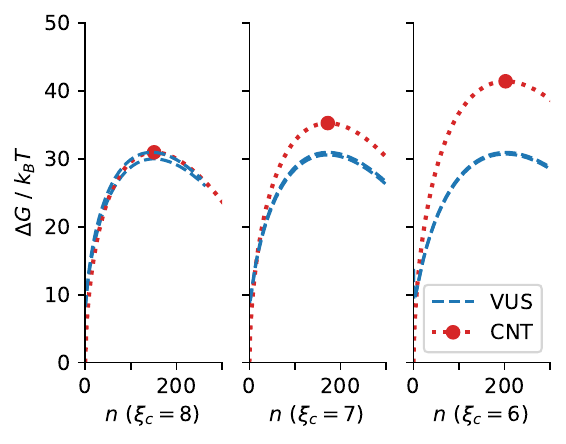}
    \caption{The free-energy barrier $\Delta G(n)/k_B T$ for the formation of a crystal nucleus of size $n$ from a fluid of nearly hard spheres (WCA) at a pressure $P \sigma^3 / k_B T= 12.0$ and temperature $k_B T/\epsilon=0.025$, for different order parameter thresholds $\xi_c$. A particle is classified as solid-like if it has at least $\xi_c$ solid-like bonds, with $\xi_c=6,7,8$. The red dots and red dotted lines represent CNT approximations, \Cref{eq:cnt-G-n}, of the barrier based on the critical nucleus size $n_c$.  The blue dashed lines are aCNT approximations of the barrier derived from Variational Umbrella Seeding (VUS) simulations. Note that two aCNT approximations are shown for each order parameter threshold $\xi_c$: one for the approximate $NPT$ and one for the rigorous $NVE$ integrator used in the hybrid Monte Carlo simulations. As they are almost indistinguishable, we do not label them separately.}
    \label{fig:wca}
\end{figure}

With our results for nearly hard spheres, we can illustrate the order parameter {independence} of Variational Umbrella Seeding. The nucleation barrier height is {nearly independent} of the order parameter threshold $\xi_c$ that is used to identify the crystal nucleus: while varying $\xi_c$ from $6$ to $8$, the nucleation barriers estimated by Variational Umbrella Seeding varies less than $1 ~k_B T$. In contrast, the critical nucleus size $n_c$ depends sensitively on the chosen order parameter. Consequently, the CNT approximation $\Delta G^c = n_c |\Delta \mu| / 2$ also relies heavily on the order parameter. For nearly hard spheres, the CNT approximation decreases from $\Delta G^c=40.8 ~k_B T$ to $\Delta G^c=30.8 ~k_B T$ when changing the order parameter from $\xi_c=6$ to $\xi_c=8$. In \Cref{fig:wca}, we illustrate the order parameter {independence} of Variational Umbrella Seeding for nearly hard spheres. The red dotted lines represent  CNT approximations of the nucleation barrier based on the critical nucleus size, whereas the dashed blue lines are the aCNT fits from the Variational Umbrella Seeding simulations. For $\xi_c=8$, these barriers are nearly identical. However, for $\xi_c=7$ and $\xi_c=6$, the critical nucleus size increases, and the CNT approximation overestimates the barrier height. Additionally, we observe that the approximate $NPT$ and the rigorous $NVE$ integrator yield almost identical results for the nucleation barrier. In \Cref{fig:wca}, we plotted the nucleation barriers from both schemes, i.e.\ one blue dashed line corresponds to the $NPT$ and one to the $NVE$ integrator. Since these lines almost exactly overlap, we conclude that these schemes give almost identical results.

\subsection{Monatomic water (mW)}
The second model we investigate is the monatomic model of water (mW).\cite{molinero2009water} Following Ref.\ \citenum{russo2014new}, we study three different temperatures: $T=\SI{215.1}{\kelvin}$, $\SI{225.0}{\kelvin}$, and $\SI{235.0}{\kelvin}$, while fixing zero pressure. The melting temperature for this model at zero pressure is $T_m = \SI{274.6}{\kelvin}$.\cite{molinero2009water} For $\SI{215.1}{\kelvin}$ and $\SI{225.0}{\kelvin}$, we use \proof{2000}{$N=2\times 10^3$} particles, and for $\SI{235.0}{\kelvin}$, we employ \proof{4000}{$N=4\times 10^3$} particles,  following Ref.\ \proof{\citenum{russo_interplay_2013}}{\citenum{russo2014new}}. The timestep is $2$ fs, and the temperature and pressure are fixed using a Nos\'e-Hoover thermostat and barostat with a relaxation time of $\SI{0.1}{\pico\second}$ and $\SI{0.2}{\pico\second}$, respectively. Following Ref.~\citenum{espinosa_seeding_2016}, we identify the crystal nucleus as follows: we first compute the averaged bond-order parameter $\bar{q}_6$ \replace{}{for each particle}.\cite{lechner2008accurate} Employing a nearest-neighbor cutoff of $\SI{3.51}{\angstrom}$, solid-like particles are identified as $\bar{q}_6 > \bar{q}_6^c$, where the threshold $\bar{q}_6^c$ depends on temperature as $\bar{q}_6^c = 0.5055 - 5.324 \times 10^{-4} ~T/\mathrm{K}$. 
\replace{}{Two solid-like particles belong to the same cluster if their distance is less than $\SI{3.51}{\angstrom}$. Finally, the largest cluster of solid-like particles is identified as the crystal nucleus.}
Based on the seeding simulations of Ref.\ \citenum{espinosa_seeding_2016}, we use the following initial estimates for the critical nucleus sizes: $\hat{n}_c=70$, $150$, and $320$  for $T=215.1$, $225.0$, and $\SI{235.0}{\kelvin}$, respectively. In our hybrid Monte Carlo scheme, we used short molecular dynamics trajectories with a  length of $\SI{0.5}{\pico\second}$ amounting to a simulation time of $\SI{400}{\nano\second}$ for each target size. We used the first $\SI{40}{\nano\second}$ for equilibration and the following  $\SI{360}{\nano\second}$ for production.
We determined an autocorrelation time of approximately $0.2$ ns, which  did not exhibit significant  temperature dependence. To compute the nucleation rate from the nucleation barrier, we use a constant kinetic prefactor $J_0 = 5\times 10^{39} \mathrm{m}^{-3} \mathrm{s}^{-1}$ because, within the temperature range we consider, it  remains approximately constant. This kinetic prefactor is simply an average of the kinetic prefactors reported in Ref.~\citenum{russo2014new}.

\begin{figure}[tbp]
    \centering
    \includegraphics{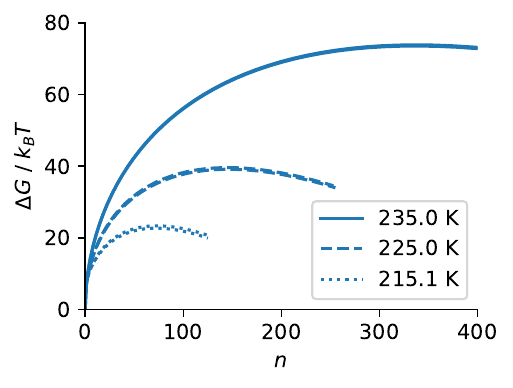}
    \caption{Free-energy barriers $\Delta G(n)/k_B T$ for the formation of a crystal nucleus of size $n$ from an mW liquid at zero pressure for different temperatures $T$. The blue lines are aCNT approximations of the barrier based on Variational Umbrella Seeding (VUS) simulations. Note that two aCNT approximations are shown for each temperature: one for the approximate $NPT$ and one for the rigorous $NVE$ integrator for the hybrid Monte Carlo simulations. As they are almost indistinguishable, we do not label them separately.}
    \label{fig:mW-barriers}
\end{figure}

\begin{figure}[thbp]
    \centering
    \includegraphics[width=\linewidth]{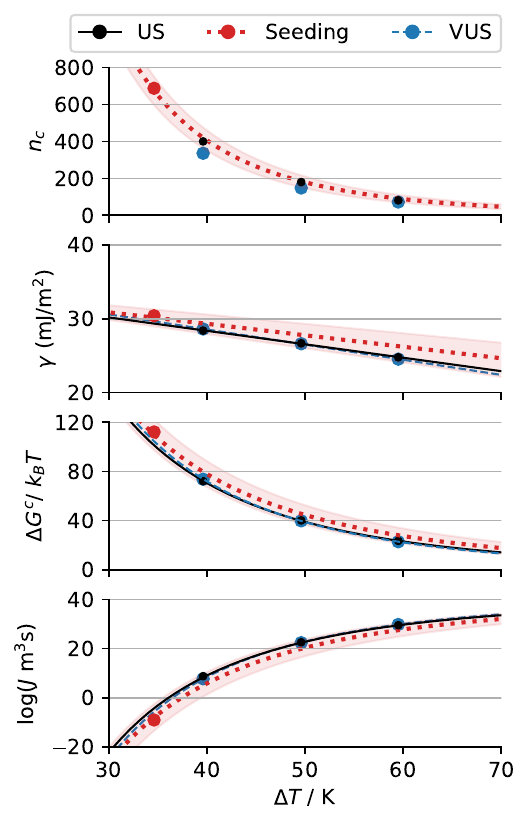}
    \caption{Critical nucleus size $n_c$, interfacial tension $\gamma$, nucleation barrier $\Delta G^c / k_B T$, and decimal logarithm of the nucleation rate $J \mathrm{m^3 s}$ as a function of supercooling $\Delta T$ at zero pressure for the mW model of water. The black dots and solid lines are Umbrella Sampling (US) results from Ref.~\citenum{russo2014new}, the red dots and red dotted lines are Seeding results from Ref.~\citenum{espinosa_seeding_2016}, and the blue dots and dashed lines are our Variational Umbrella Seeding (VUS) results.
    }
    \label{fig:mW}
\end{figure}


\begin{figure*}[thbp]
    \centering
    \includegraphics[width=\linewidth]{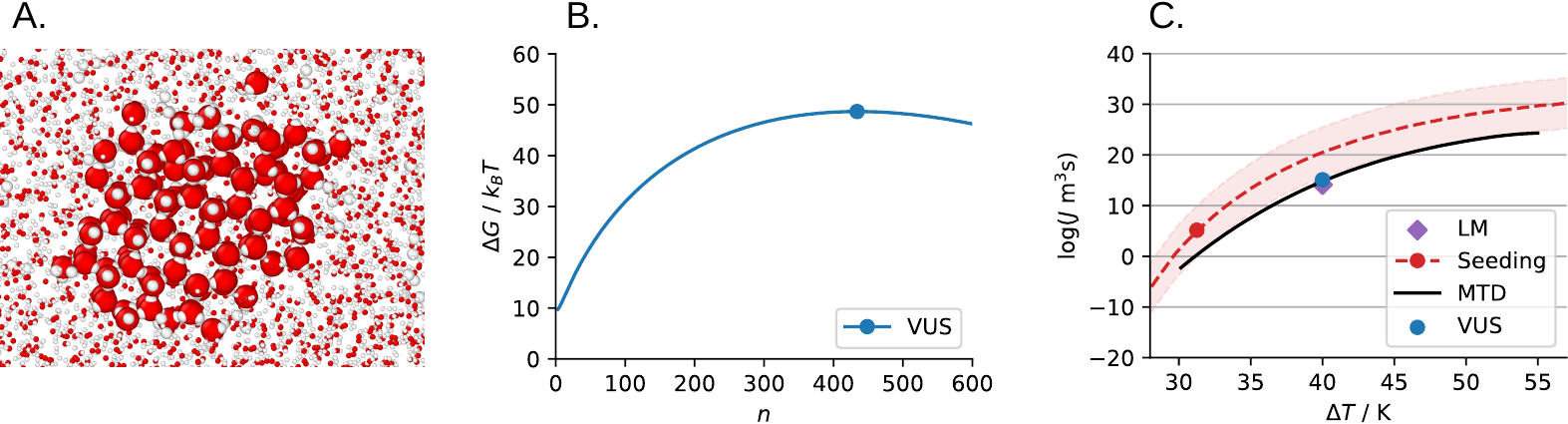}
    \caption{
    Crystal nucleation of the TIP4P/ICE model of water at $P=\SI{1}{bar}$.
    A. Cut-through image of a crystal nucleus of size $n\approx 240$ at $T=\SI{230}{K}$. The nucleus is identified as the largest cluster of particles that have an averaged bond-order parameter $\bar{q}_6 > 0.378$. The surrounding particles are reduced in size for clarity. 
    B. Free-energy barrier $\Delta G(n)/k_B T$ for the formation of a crystal nucleus of size $n$ 
    at $T=\SI{230}{\kelvin}$. The blue line is the aCNT approximation of the barrier based on our Variational Umbrella Seeding (VUS) simulations. The blue dot marks the top of the barrier.
    C. Decimal logarithm of the nucleation rate $J \mathrm{m^3 s}$ as a function of supercooling $\Delta T$.
    The red dashed line and red shaded region are Seeding results from Ref.\ \citenum{espinosa2016time}, while the black solid line is from Metadynamics (MTD) simulations from Ref.\ \citenum{niu2019temperature}. The purple diamond is from Lattice Mold (LM) simulations of Ref.\ \citenum{sanchez2022homogeneous}. The blue dot is the Variational Umbrella Seeding (VUS) result of this work.
    }
    \label{fig:tip4p}
\end{figure*}

In \Cref{fig:mW-barriers}, we plot our Variational Umbrella Seeding results for the nucleation barriers of mW. We present both the  $NPT$ and  $NVE$ integrator results simultaneously. Again, we observe that these schemes yield nearly identical results for the shape and height of the nucleation barrier.

In \Cref{fig:mW}, we show the critical nucleus size $n_c$, interfacial tension $\gamma$, barrier height $\Delta G^c$, and nucleation rate $J$ as a function of supercooling $\Delta T = T_m - T$ for mW. We compare our Variational Umbrella Seeding (VUS) results with previous Umbrella Sampling (US) results by \citet{russo2014new} and Seeding results by \citet{espinosa_seeding_2016} Strictly speaking, the Seeding simulations\cite{espinosa_seeding_2016} were performed at $\SI{1}{\bar}$, whereas the Umbrella Sampling simulations\cite{russo2014new} and our Variational Umbrella Seeding simulations were performed at zero pressure. Since the slope of the melting curve of ice Ih is on the order of  $-(100-200)$ bar/K,\cite{vega2005melting} from 1 bar to 0 bar the melting temperature changes less than $0.01$ K. In the following analysis, we ignore this very minor difference.

At a supercooling of $\Delta T = 40 \mathrm{K}$, one can see that the critical nucleus size $n_c$ reported in Ref.\ \citenum{russo2014new} (black dot) differs slightly from our value (blue dot). As discussed before, this is simply a result of using different criteria to estimate the nucleus size.
Using the CNT approximation, these different critical nucleus sizes automatically lead to different nucleation barriers and nucleation rates. The red-shaded regions in \Cref{fig:mW} show the associated uncertainty in the nucleation barriers and rates. \citet{espinosa_seeding_2016} estimate that in this case, the dependence of Seeding on the choice of order parameter leads to an uncertainty of about four orders of magnitude in the nucleation rate.

As observed in the case of nearly hard spheres, Variational Umbrella Seeding exhibits no such order parameter dependence. In \Cref{fig:mW}, we plot the Variational Umbrella Seeding estimates of the nucleation barrier and nucleation rate with blue dots and dashed blue lines. Additionally, we  plot the Umbrella Sampling results of Ref. \citenum{russo2014new} in this figure using black dots and solid lines. Comparing Umbrella Sampling, Seeding, and Variational Umbrella Seeding, we see that they all show good agreement. As previously found, the mislabeling criterion is a very reasonable choice of order parameter for Seeding simulations. Notably, our Variational Umbrella Seeding results are almost indistinguishable from Umbrella Sampling. 


\subsection{TIP4P/ICE}

The final model we investigate is the TIP4P/ICE model of water at $T=230 K$ and $P=\SI{1}{\bar}$, for which the melting temperature is $T_m = \SI{270}{\kelvin}$.\cite{espinosa2016time} We simulate \proof{5881}{$N=6\times 10^3$} molecules using a timestep of $2$ fs, while both the Nos\'e-Hoover thermostat and barostat have a relaxation time of $2$ ps. Pair interactions are truncated at $\SI{9.0}{\angstrom}$. For long-range Coulombic interactions, we use the pppm/tip4p particle–particle particle-mesh (PPPM) solver in LAMMPS\cite{hockney2021computer} with a relative error in forces of $10^{-5}$. Long range corrections to energy and pressure were included for the Lennard-Jones part of the potential.  In addition, we employ the Shake algorithm\cite{ryckaert1977numerical} to constrain the O-H bond lengths and H-O-H angles. As the Shake algorithm is not reversible, the $NVE$ integrator does not guarantee detailed balance. Hence, we solely performed simulations using the approximate $NPT$ integrator. Again, we use the averaged bond-order parameter $\bar{q}_6$ as an order parameter\replace{,}{. We compute $\bar{q}_6$ for each oxygen atom using only the positions of  the oxygen atoms,} with a \replace{}{nearest-neighbor} cutoff distance of $\SI{3.51}{\angstrom}$. By employing a linear extrapolation of the thresholds in Ref.\ \citenum{espinosa2016time}, we selected the threshold $\bar{q}_6^c = 0.378$\replace{.}{, so an oxygen atom is considered solid-like if $\bar{q}_6 > 0.378$. Two solid-like particles belong to the same cluster if their distance is less than $\SI{3.51}{\angstrom}$. Finally, the largest cluster of solid-like particles is identified as the crystal nucleus.} Based on the Seeding simulations of Ref.\ \citenum{espinosa2016time}, we use $\hat{n}_c=240$ as an initial estimate for the critical nucleus size. We used short molecular dynamics trajectories with a  length of $\SI{20}{\pico\second}$ amounting to a simulation time of $\SI{650}{\nano\second}$ for each target size. We used the first $\SI{200}{\nano\second}$ for equilibration and the following $\SI{450}{\nano\second}$ for production. We found an autocorrelation time of approximately $\SI{20}{\nano\second}$. 


In \Cref{fig:tip4p}, we show our results for TIP4P/ICE. In \Cref{fig:tip4p}A, we visualize a crystal nucleus of size $n\approx 240$, where we can see clearly its hexagonal structure. In \Cref{fig:tip4p}B, we show the nucleation barrier $\Delta G(n)$ as a function of nucleus size $n$ for $T=\SI{230}{\kelvin}$. The blue line is the aCNT fit of the nucleation barrier. 
\replace{W}{Most importantly, w}e find a barrier height of $50.4~k_B T$ with a statistical error of around $6 ~k_B T$. Previous Metadynamics simulations\cite{niu2019temperature} found a barrier height of $52.8~k_B T$ with a similar statistical error of around $6~k_B T$. Therefore, we conclude that our Variational Umbrella Seeding result \replace{}{for the barrier height} agrees well with the Metadynamics simulations.

In \Cref{fig:tip4p}C, we compare the nucleation rate $J$ of TIP4P/ICE as a function of supercooling, obtained with Seeding \cite{espinosa2016time}, Metadynamics \cite{niu2019temperature}, Lattice Mold \cite{sanchez2022homogeneous,tejedormold}, and our Variational Umbrella Seeding. To compute the nucleation rate from our nucleation barrier, we use the same kinetic prefactor of $J_0 = 10^{37} m^{-3} s^{-1}$ as was used in previous Seeding simulations.\cite{espinosa2014homogeneous} Seeding has an uncertainty of around 5 orders of magnitude represented by the red shaded region. We observe  that our result for the nucleation rate with Variational Umbrella Seeding (blue dot) agrees very well with previous results from Metadynamics \cite{niu2019temperature} and Lattice Mold \cite{sanchez2022homogeneous} simulations.




\subsection{Summary of barriers}
In \Cref{tab:wca}, we summarize our Variational Umbrella Seeding results for the nucleation barriers. Additionally, \Cref{tab:wca} presents  reference results obtained with Umbrella Sampling for WCA \cite{filion_simulation_2011} and mW \cite{russo2014new} or metadynamics for TIP4P/ICE.\cite{niu2019temperature} The barrier height varies from $\Delta G^c=\proof{21}{22} ~k_B T$ to $72 ~k_B T$, and the critical nucleus size ranges from $n_c=72$ to $453$. Across this wide variety of barriers and models, we observe that Variational Umbrella Seeding agrees well with the reference results within $2-3 ~k_B T$ in all cases. In general, we obtain slightly lower values for the barrier height.
Both the $NPT$ and the $NVE$ integrator schemes give \replace{accurate}{consistent} results, indicating that they can both be used to obtain approximations of the barrier height.

\setlength{\tabcolsep}{7pt} 
\renewcommand{\arraystretch}{1.3} 
\begin{table*}[!t]
    \centering
    \begin{tabular}{r|l|r|c|r|r|r|r|r|r|r}
    Label & Model & $T$     &Order parameter& $N$ & $n_c$ & $|\Delta\mu|$ & \multicolumn{4}{c}{$\Delta G^c ~/~ k_B T$} \\ 
   & & & & & & & $n_c |\Delta\mu|/2$ & \multicolumn{2}{c|}{\textbf{VUS}}   & Reference \\ 
     & &&&&& &  &  $NPT$  & $NVE$  & \\ \hline 
     I & WCA     & $0.025 ~\epsilon/k_B $      &$\xi_c=6$  & \proof{$2\times 10^4$}{$8\times 10^3$}      &$203$ &0.41\cite{filion_simulation_2011} & $41.6$ & \proof{30.2}{$30.6$}	   &\proof{30.7}{$31.1$} &{$32.5$} \cite{filion_simulation_2011}  \\ 
     II & WCA    &  $0.025 ~\epsilon/k_B $      &$\xi_c=7$  & \proof{$2\times 10^4$}{$8\times 10^3$}       &$170$   &0.41\cite{filion_simulation_2011} & $34.9$ & \proof{30.3}{$30.6$}	   &\proof{30.5}{$30.8$}  &{$32.5$} \cite{filion_simulation_2011} 	\\
     III & WCA    &  $0.025 ~\epsilon/k_B $      &$\xi_c=8$  & \proof{$2\times 10^4$}{$8\times 10^3$}       &$147$    &0.41\cite{filion_simulation_2011} & $30.1$ & \proof{29.8}{$29.9$}	   &\proof{30.8}{$31.1$} &{$32.5$} \cite{filion_simulation_2011}  \\ \hline
     IV & mW      & $\SI{215.1}{\kelvin}$&  $\bar{q}_6^c=0.391$   & \proof{2000}{$2\times 10^3$}      & $72$  & 0.62 \cite{espinosa2014homogeneous}  & $22.3$ & \proof{21.0}{$22.4$}	   &\proof{21.7}{$23.1$}   &{$23.5$}\cite{russo2014new} \\ 
     V & mW      & $\SI{225.0}{\kelvin}$&  $\bar{q}_6^c=0.386$  & \proof{2000}{$2\times 10^3$}     & $148$ & 0.50 \cite{espinosa2014homogeneous} & $37.0$ & \proof{38.8}{$40.0$}	   &\proof{38.5}{$39.5$}   &{$40.1$} \cite{russo2014new} \\
     VI & mW      & $\SI{235.0}{\kelvin}$&  $\bar{q}_6^c=0.380$  & \proof{4000}{$4\times 10^3$}     & $336$   &0.38 \cite{espinosa2014homogeneous} & $63.8$ &\proof{71.8}{$73.4$}	   &\proof{72.6}{$73.7$}   &$72.0$\cite{russo2014new} \\ \hline
    VII & TIP4P/ICE    & 230.0 K& $\bar{q}_6^c=0.378$   & \proof{5881}{$6\times 10^3$}     & $453$ &0.32 \cite{espinosa2016time} & $72.5$ & $50.4$   & -     &$52.8$ \cite{niu2019temperature} \\ 
    \end{tabular}

    \caption{Results from Variational Umbrella Seeding (VUS) for the nucleation barriers $\Delta G^c / k_B T$ of nearly hard spheres (WCA), a monatomic model of water (mW), and the TIP4P/ICE model of water are compared with reference results from Umbrella Sampling for WCA\cite{filion_simulation_2011} and mW, \cite{russo2014new} or metadynamics for TIP4P/ICE.\cite{niu2019temperature} For WCA,  the pressure $P \sigma^3 /k_B T = 12.0$ and the temperature $T=0.025 ~\epsilon / k_B$ are fixed, while varying the order parameter for identifying the crystal nucleus, leading to different critical nucleus sizes $n_c$ and therefore different Classical Nucleation Theory (CNT) approximations $\Delta G_{\mathrm{CNT}}^c = n_c|\Delta\mu|/2$. For mW and TIP4P/ICE,  the temperature $T$ is varied, but  the order parameter threshold is fixed using the mislabeling technique. The pressure is $0$ and $1$ bar, respectively, for mW and TIP4P/ICE. \replace{}{The system size $N$ refers to the total number of particles in the simulation for WCA and mW, and to the total number of molecules for TIP4P/ICE.} The barriers $\Delta G^c / k_B T$ shown in the Variational Umbrella Seeding (VUS) and Reference columns were all measured in the $NPT$ ensemble. For Variational Umbrella Seeding, we explore two different choices for the integrator in our hybrid Monte Carlo / molecular dynamics scheme: the approximate $NPT$ and the rigorous $NVE$ integrator. All values shown in the columns from `$|\Delta\mu|$' to `Reference' are in units of $k_B T$. The statistical error in $n_c$ is around $5$ particles for WCA and mW, and around $60$ particles for TIP4P/ICE. \replace{}{The critical nucleus size $n_c$ is obtained from the maximum of the adjusted Classical Nucleation Theory (aCNT) fit of the barrier.} The statistical error in our VUS barriers is around $0.5~k_B T$ for WCA and mW, and around $6 ~k_B T$ for TIP4P/ICE.
    }
    \label{tab:wca}
\end{table*}

\subsection{Comparison with existing methods}
\label{sec:comparison}
Now that we have demonstrated that Variational Umbrella Seeding successfully  reproduces  nucleation barriers obtained by Umbrella Sampling and metadynamics, it is useful to discuss the advantages of Variational Umbrella Seeding compared to these and other existing methods. In short, we argue that Variational Umbrella Seeding is faster than rigorous methods, while being more {accurate} than Seeding.

Firstly, we have shown that Variational Umbrella Seeding is \replace{independent of}{less sensitive to} the choice of order parameter. We have already discussed how this improves upon the original Seeding approach, but this argument also applies to other methods such as the nucleus size pinning method\cite{sharma2018nucleus} and the persistent embryo approach,\cite{sun2018overcoming} since these methods also rely on the CNT approximation, \Cref{eq:cnt-n}. 

Secondly, using the aCNT \Cref{eq:acnt} for the nucleation barrier reduces the problem to fitting three parameters. In standard Umbrella Sampling, one parameter must be estimated per window to stitch the windows together. Because standard Umbrella Sampling requires overlapping windows, the number of windows quickly increases with the size of the critical nucleus. In Variational Umbrella Seeding, there is no need for overlap between different windows as there is no need to stitch these together. Therefore, Variational Umbrella Seeding only requires four windows, independent of the critical nucleus size, much less than standard Umbrella Sampling. This advantage of Variational Umbrella Seeding with respect to standard Umbrella Sampling will be more pronounced for smaller supersaturations and larger critical nucleus sizes. \replace{}{To further quantify this advantage, we can  estimate the number of windows required for standard Umbrella Sampling. Typically, the first $10~k_B T$ of a nucleation barrier can  be explored with an unbiased simulation. With our choice of spring constants (see \Cref{sec:idealized}),  windows should be positioned in $2~k_B T$ intervals to ensure ample overlap between them. Hence, the approximate number of windows for standard Umbrella Sampling is  $((\Delta G^c ~/~ k_B T) - 10)/2$. For our barriers, this translates to 11 windows for WCA, 7, 15, and 31 for mW at temperatures $T=\SI{215.1}{\kelvin}$, $\SI{225.0}{\kelvin}$, and $\SI{235.0}{\kelvin}$, respectively, and roughly 21 for TIP4P/ICE. Consequently, using Variational Umbrella Seeding,  the number of windows can be reduced by a factor $2$ to $8$ compared to standard Umbrella Sampling, with this reduction factor increasing further for higher barriers.}


We expect that the choice of order parameter will also influence the efficiency of the CNT-US method by \citet{russo2014new} and the method of \citet{mccarty2015variationally} Both of these methods rely on a bias potential of the CNT form, \Cref{eq:cnt}. We have seen, for example in \Cref{fig:hs}, that this CNT approximation can be less appropriate, especially for the initial part of the barrier. 
This may be problematic in the initial part of the barrier where it is very steep. 
In these higher gradient regions, stronger biases are needed, resulting in lower acceptance ratios for Monte Carlo schemes. This decreases the efficiency of biased simulations.
In contrast, in Variational Umbrella Seeding,  there is no need to perform biased simulations of the initial part of the barrier. The metastable liquid is simulated without any bias, and all the biased simulations are performed at higher nucleus sizes, where the gradient of the nucleation barrier is less steep.

\replace{}{Although we have observed that Variational Umbrella Seeding is significantly less sensitive to the choice of order parameter than Seeding, it is not entirely independent of this choice. Similar to standard Umbrella Sampling and metadynamics calculations, the `collective variable' - i.e. the coordinate biased with the bias potential - should serve as a reasonable `reaction coordinate'. If this is not the case, then calculations of the barrier height and nucleation rate may become unreliable. In this work, the collective variable is the size of the nucleus. As discussed in Refs.\ \citenum{agarwal2014solute,peters2006using},  the quality of a collective variable can be assessed through committor analyses. Alternatively, accuracy can be tested by comparing nucleation rates with brute-force measurements. Empirical evidence for nucleation of hard spheres,\cite{filion_crystal_2010,filion_simulation_2011,gispen2023brute,gispen2024finding} mW,\cite{espinosa_seeding_2016} Lennard-Jones,\cite{espinosa_seeding_2016,bulutoglu2023comprehensive} and NaCl\cite{espinosa_seeding_2016}, suggests that the size of the nucleus serves as a reasonable collective variable for these systems. However, caution should always be taken to ensure that the collective variable does not miss alternative reaction pathways. 
For example, to examine the competition between different crystal polymorphs during nucleation, it is necessary to adapt the collective variable to each individual polymorph.\cite{agarwal2014solute,sanchez-burgos_fcc_2021,sanchez-burgos_parasitic_2021,garaizar_alternating_2022,gispen2022kinetic,gispen2023crystal} As another example, to examine the competition between amorphous and crystalline solid phases during nucleation, it may be necessary to employ two collective variables: the size of the largest amorphous cluster and the size of the largest crystalline cluster.\cite{agarwal2014solute}
}



\replace{}{Similar to the nucleus size pinning method,\cite{sharma2018nucleus} the persistent embryo method,\cite{sun2018overcoming} and to $NVT$-Seeding,\cite{rosales2020seeding} an advantage of Variational Umbrella Seeding over Seeding is that the nuclei can equilibrate their interface with the surrounding parent phase. Consequently, as explained in Ref. \citenum{sharma2018nucleus}, nuclei have the potential to alter their shape during a simulation. It is important to note that while Seeding does not inherently assume a spherical shape, as seeds of various shapes can be used, Variational Umbrella Seeding is generally less sensitive to the preparation method of the initial configuration than Seeding.}

An advantage with respect to metadynamics is that there is no need to compute bias forces in Variational Umbrella Seeding. This advantage has several aspects. Firstly, this makes the choice of order parameter much more flexible. In Variational Umbrella Seeding, the order parameter does not need to be differentiable. For instance, when the number of particles in the nucleus is employed as an order parameter, the order parameter takes only integer values. Therefore, the bias potential is not a continuous function of the order parameter and one cannot compute the derivative that is needed to evaluate the force.
Even when order parameters are differentiable, it can be inconvenient or computationally expensive to calculate their derivatives. Secondly,  order parameters do not have to be computed for every simulation timestep. Depending on the  complexity of the model and the order parameter, this can lead to  a significant speedup. Thirdly, Variational Umbrella Seeding is compatible with conventional molecular dynamics or Monte Carlo codes by coupling them to an external hybrid Monte Carlo scheme. There is no need to modify the molecular dynamics or Monte Carlo codes themselves, eliminating the need to `patch' a program with additional software such as  PLUMED.\cite{bonomi_promoting_2019}

For a quantitative comparison, we can compare the simulation time required to estimate the nucleation rate of TIP4P/ICE at $\SI{230}{\kelvin}$. Seeding simulations required around $1.5\times 10^5$ central processing unit (CPU) hours.\cite{espinosa2014homogeneous} With Lattice Mold simulations, this increased to around $1\times 10^6$ CPU hours.\cite{respinosa_lattice_2016} Forward flux sampling calculations needed approximately $2\times 10^7$ CPU hours.\cite{haji2015direct} For our Variational Umbrella Seeding simulations we used around $2\times 10^5$ CPU hours. This comparison suggests that Variational Umbrella Seeding is more efficient than Lattice Mold, significantly more efficient than forward flux sampling, and only slightly less efficient than Seeding.


\section{Conclusion}

In conclusion, we introduced Variational Umbrella Seeding, a novel computational technique for estimating nucleation barriers. The theoretical basis of this method is  adjusted classical nucleation theory. Hybrid molecular dynamics - Monte Carlo simulations are used to obtain segments of the nucleation barrier. We then fit the free parameters in the adjusted classical nucleation theory using a variational approach. Our results demonstrate  good agreement with previous methods for estimating crystal nucleation barriers of nearly hard spheres (WCA), monatomic water (mW), and the TIP4P/ICE model of water. The  nucleation barrier values ranged from $20 ~k_B T$ to $72 ~k_B T$. Variational Umbrella Seeding  matched all these previous results within $3 ~k_B T$, which is the typical uncertainty range for these calculations. 

Given its low computational cost, we believe that Variational Umbrella Seeding can serve as a valuable tool for investigating nucleation rates across  wide ranges of temperatures, pressures, and particle interactions. Moreover, we anticipate that this technique can be extended to explore other nucleation processes such as condensation, cavitation, and melting.

\section*{Supplementary Material}
\replace{Please see t}{T}he supplementary material \replace{for}{contains:} the code used to generate and analyze the results of this paper\replace{.}{; initial configurations used for the simulations; nucleus size distributions from the unbiased liquid phase; nucleus size distributions from the biased simulations.} 

\begin{acknowledgments}
M.D. and W.G. acknowledge funding from the European Research Council (ERC) under the European Union’s Horizon 2020 research and innovation programme (Grant
agreement No. ERC-2019-ADG 884902 SoftML). J.R.E. acknowledges funding from the Ramon Y Cajal fellowship (RYC2021-030937-I) and the Spanish National Agency for Research (PID2022-136919-NAC33). E.S and C.V  acknowledge funding from MEC by grant 
PID2022-136919NB-C31.
\vspace{1em}
\end{acknowledgments}

\section*{Data Availability Statement}
The data that supports the findings of this study are available within the article and its supplementary material.

%


\appendix


\section{Code}
The code used to generate and analyze the results of this paper is freely available at 
\href{https://github.com/MarjoleinDijkstraGroupUU/VariationalUmbrellaSeeding}{\url{https://github.com/MarjoleinDijkstraGroupUU/VariationalUmbrellaSeeding}} and in the supplementary material of this paper. The implementation is based on the hybrid Monte Carlo code by \citet{guo2018hybrid} and uses the LAMMPS code for molecular dynamics\cite{plimpton_fast_1995} and the freud library for calculating bond order parameters.\cite{freud2020}




\section{Barriers and fitting parameters}
In \Cref{tab:params}, we present the values of the parameters of the aCNT expression,  \Cref{eq:acnt2}, that we use to fit the nucleation barriers of WCA, mW, and TIP4P/ICE. \replace{}{In the same table, we also present the nucleus size $n_0$ where the unbiased initial part of the barrier is `glued' to the aCNT barrier, see \Cref{eq:glue-g0}.}

\setlength{\tabcolsep}{7pt} 
\renewcommand{\arraystretch}{1.3} 
\begin{table*}[!hbtp]
    \centering
    \begin{tabular}{r|l|r|r|r|r|r|r|r|r} 
    
  Label     &Model & $|\Delta\mu|$ &\multicolumn{2}{c|}{$g_2$} & \multicolumn{2}{c|}{$g_1$} & \multicolumn{2}{c|}{$g_0$} & $\replace{}{n_0}$ \\ 
  &&& $NPT$ & $NVE$ & $NPT$ & $NVE$ & $NPT$ & $NVE$ & \\ \hline
     I      &WCA & 0.41\cite{filion_simulation_2011} &4.32 & 4.31 & -8.35 & -8.21 & \proof{13.46}{13.79} &\proof{13.22}{13.56} &  \proof{6}{5} \\ 
     II     &WCA & 0.41\cite{filion_simulation_2011} &3.92 & 3.90 & -5.56 &-5.33 & \proof{10.51}{10.78} &\proof{10.21}{10.50}	& \proof{4}{3} \\
     III    &WCA & 0.41\cite{filion_simulation_2011} & 3.64 &3.50 & -3.91 &-2.75  & \proof{9.39}{9.51} &\proof{7.99}{8.26} & \proof{3}{2} \\ \hline
     IV     &mW & 0.62 \cite{espinosa2014homogeneous} & 4.29 &4.25 & -3.47 &-2.98 & \proof{6.15}{7.54}	& \proof{5.43}{6.86} & \proof{5}{4}\\ 
     V      &mW & 0.50 \cite{espinosa2014homogeneous}& 3.95 &3.99 & -0.06&-0.45 & \proof{2.22}{3.42}& \proof{2.74}{3.90}	 & \proof{4}{3}\\
     VI     &mW &0.38 \cite{espinosa2014homogeneous}& 3.77 &3.77 & 2.69  &2.75 & \proof{-1.23}{0.32}& \proof{-1.32}{0.23}	 & 3\\ \hline
VII         &TIP4P/ICE &0.32 \cite{espinosa2016time} & 4.32 & - & -9.89 & - & 16.02 & -  & \replace{}{5}  \\ 
    \end{tabular}
    \caption{Fitting parameters of the aCNT expression given by \Cref{eq:acnt2} for the nucleation barriers of WCA, mW, and TIP4P/ICE. The labels correspond to the labels  shown in \Cref{tab:wca}. The values for the supersaturation $|\Delta\mu|$ are from Refs. \citenum{filion_simulation_2011}, \citenum{espinosa2014homogeneous}, and \citenum{espinosa2016time} for WCA, mW, and TIP4P/ICE, respectively. \replace{All values}{The fitting parameters $g_i$} are shown in units of $k_B T$. \replace{}{The last column shows the nucleus size $n_0$ where the unbiased initial part of the barrier is `glued' to the aCNT barrier, see \Cref{eq:glue-g0}.} }
    \label{tab:params}
\end{table*}

\setlength{\tabcolsep}{7pt} 
\renewcommand{\arraystretch}{1.3} 
\begin{table*}[!hbtp]
    \centering
    \begin{tabular}{ r|l|r|r|r|r|r} 

  Label     &Model & $\hat{n}_c$ & $k(\hat{n}_c/4)$ & $k(\hat{n}_c/2)$ & $k(3 \hat{n}_c/4)$ & $k(\hat{n}_c)$ \\ \hline
     I      &WCA        & 185   & 0.0284 & 0.0113 & 0.0065 & 0.0044   \\ 
     II     &WCA        & 155   & 0.0333 & 0.0132 & 0.0078 & 0.0053	 \\
     III    &WCA        & \proof{103}{130}   & 0.0409 & 0.0159 & 0.0093 & 0.0063 \\ \hline
     IV     &mW         &  70   & 0.1144 & 0.0437 & 0.0257 & 0.0173 \\ 
     V      &mW         & 150   & 0.0416 & 0.0162 & 0.0095 & 0.0064	 \\
     VI     &mW         & 320   & 0.0147 & 0.0058 & 0.0034 & 0.0023 \\ \hline
VII         &TIP4P/ICE  & 240   & 0.0169 & 0.0067 & 0.0039 & 0.0027  \\ 
    \end{tabular}
    \caption{Bias potentials employed for the nucleation barriers of WCA, mW, and TIP4P/ICE. The labels correspond to the labels shown in \Cref{tab:wca}. The initial estimate $\hat{n}_c$ of the critical nucleus size determines the target sizes  $\hat{n} = \hat{n}_c/4$, $\hat{n}_c/2$, $3 \hat{n}_c/4$, $\hat{n}_c$ as well as the spring constants $k$ of the bias potentials, see \Cref{sec:idealized}.
    The spring constants $k$ are shown in units of $k_B T$. }
    \label{tab:bias-potentials}
\end{table*}



\section{Confidence interval}
In addition to obtaining a maximum-likelihood estimate for the nucleation barrier $\Delta G^c$, we can compute a confidence interval for the critical nucleus size and the barrier height  using the likelihood function, \Cref{eq:logL}.
To do this, we follow the procedure outlined in Ref.\ \citenum{venzon_method_1988}. To determine an approximate $95\%$ confidence interval for the nucleation barrier $\Delta G^c$, we vary the parameters $\hat{g}_2+dg_2$ and $\hat{g}_1 + dg_1$ and compute the likelihood $L$ relative to the likelihood of the maximum likelihood estimate, i.e.\ $ L_{\mathrm{rel}} = L(\hat{g}_2+dg_2, \hat{g}_1+dg_1) /  L(\hat{g}_2, \hat{g}_1)$. The values $dg_2,dg_1$ for which this relative likelihood $L_{\mathrm{rel}} = 0.05$ represent the boundaries of the $95\%$ confidence region. Note that this criterion is specific to our case of two parameters, and a different threshold should be used to convert the relative likehood to a confidence region if a different number of parameters is estimated. Once the boundaries of the confidence region are estimated, we can determine the critical nucleus size $n_c$ and barrier height $\Delta G^c$ for all values of $g_2$ and $g_1$ within the $95\%$ confidence region. In this way, we also obtain a $95\%$ confidence interval for $n_c$ and $\Delta G^c$. Although these confidence intervals should always be treated with care, they can provide  insights into the minimum simulation time  needed to fit the aCNT parameters.\\

\section{Optimizing fitting strategies with an idealized nucleation barrier}
\label{sec:idealized}
To optimize our fitting strategy, we performed tests with an idealized nucleation barrier. To be more precise, we assumed an idealized nucleation barrier $\Delta G(n)$ that is exactly described by the aCNT with exactly known parameter values $g_i$. That is, we selected values for $g_i$ and constructed $\Delta G(n)$ accordingly.
Recall that both the nucleation barrier and the external bias potential $U_{\mathrm{ext}}$ induce a probability distribution $p(n)$ for the nucleus size, given by
\begin{equation*}
    p(n) = \frac{1}{Z}\exp[- (\Delta G_{\mathrm{aCNT}} (n) +  U_{\mathrm{ext}} (n)) /k_B T].
\end{equation*}
Therefore, we can numerically generate independent samples from this probability distribution, akin to conducting  a biased simulation. Subsequently, we employed our maximum-likelihood approach to test how well we can recover our idealized nucleation barrier using a limited  number of independent samples.

Constructing and fitting an idealized nucleation barrier in this way is relatively quick. We iteratively constructed and fitted nucleation barriers of varying heights  using different fitting strategies. In this manner, we optimized the fitting strategy that we used for our simulations.


Firstly, we roughly estimate the critical nucleus size $\hat{n}_c$. This can be efficiently obtained  using the seeding technique or estimated from previous literature. We use four different windows, positioned at target sizes $\hat{n} = \hat{n}_c/4$, $\hat{n}_c/2$, $3 \hat{n}_c/4$, and $\hat{n}_c$. From the critical nucleus size estimate $\hat{n}_c$, we can calculate a first estimate of the interfacial tension $\gamma$. This is enough to have an idea of the nucleation barrier using \Cref{eq:cnt-G-n}. From \Cref{eq:cnt-G-n}, we approximate the local curvature of the nucleation barrier. In order to constrain the nucleus size, the spring constant should be larger than the local curvature. To be on the safe side, the spring constant $k$ of the bias potential is chosen to be six times\cite{kastner2005bridging} the local curvature i.e.\
\begin{equation*}
    k(\hat{n}) = 6 \left. \frac{d^2\Delta G_{\mathrm{CNT}}}{ d n^2} \right\rvert_{\hat{n}}.
\end{equation*}
This means that a different spring constant $k$ is chosen for each target size $\hat{n}$. 
The idea to base the spring constant on the curvature is from Ref.~\citenum{kastner2005bridging}. \replace{}{In \Cref{tab:bias-potentials}, we present the values of the spring constants $k$ that we use for the bias potentials in our simulations.}

%

\end{document}